\documentclass[conference]{IEEEtran}
\IEEEoverridecommandlockouts
\usepackage{cite}
\usepackage{amsmath,amssymb,amsfonts}
\usepackage{algorithmic}
\usepackage{graphicx}
\usepackage{textcomp}
\usepackage{xcolor}
\usepackage{tikz}
\begin{document}
\title{A Novel Deep Learning Architecture for Decoding Imagined Speech from EEG}

\author{\IEEEauthorblockN{Jerrin Thomas Panachakel}
\IEEEauthorblockA{\textit{Indian Institute of Science} \\
Bangalore, India \\
jerrinp@iisc.ac.in}
\and
\IEEEauthorblockN{A.G. Ramakrishnan}
\IEEEauthorblockA{\textit{Indian Institute of Science} \\
Bangalore, India \\
agr@iisc.ac.in}
\and
\IEEEauthorblockN{T.V. Ananthapadmanabha}
\IEEEauthorblockA{\textit{Voice and Speech Systems} \\
Bangalore, India\\
tva.blr@gmail.com}}

\maketitle

\begin{abstract}
The recent advances in the field of deep learning have not been fully utilised for decoding imagined speech primarily because of the unavailability of sufficient training samples to train a deep network. In this paper, we present a novel architecture that employs deep neural network (DNN) for classifying the words ``in'' and ``cooperate'' from the corresponding EEG signals in the ASU imagined speech dataset. Nine EEG channels, which best capture the underlying cortical activity, are chosen using common spatial pattern (CSP) and are treated as independent data vectors. Discrete wavelet transform (DWT) is used for feature extraction. To the best of our knowledge, so far DNN has not been employed as a classifier in decoding imagined speech. Treating the selected EEG channels corresponding to each imagined word as independent data vectors helps in providing sufficient number of samples to train a DNN. For each test trial, the final class label is obtained by applying a majority voting on the classification results of the individual channels considered in the trial. We have achieved accuracies comparable to the state-of-the-art results. The results can be further improved by using a higher-density EEG acquisition system in conjunction with other deep learning techniques such as long short-term memory.
\end{abstract}


\section{\label{sec:1} Introduction}

A strong motivation to work on imagined speech arises in the context of speech disability. Speech is one of the most basic and natural form of communication acquired by individuals, even  by the illiterates. Though exact statistics are not available, it is estimated that more than 70 million people around the world have speech disability. Speech disability due to complete paralysis prevents people from communicating with other in any modality. Complete paralysis may be a congenital disorder or an acquired one, which may be due to disease or accident. A complete recovery in these cases is rare and the cost of treatment is usually beyond the reach of many. Also, these conditions take a toll on the mental health of both the affected person and his relatives, reducing their efficiency and economic output. It will greatly help the affected person if, by some means, we are able to decode his/her thoughts, commonly referred to as ``imagined speech''.

The interest in imagined speech dates back to the days of Hans Berger who invented electroencephalogram (EEG) as a tool for synthetic telepathy \cite{la1999history}. Although it is almost a century since the first EEG recording, the success in decoding imagined speech from EEG signals is rather limited. One of the major reasons being the very low signal-to-noise ratio (SNR) of EEG signals.

The potential of the recent developments in the field of machine learning, such as deep neural networks (DNN) has not been exploited to its full potential in the field of decoding imagined speech, since such techniques require a huge amount of training data. In this paper, we select those EEG channels that best represent the underlying cortical activity of each imagined word by using CSP. The EEG channels so selected for each imagined word is considered as an independent input signal, thus providing more training data. This is in contrast to the earlier approaches concatenating the features to form a single feature vector.

The architecture includes a CSP based channel selection stage, discrete wavelet transform (DWT) based feature extraction stage, a classification stage consisting of a DNN with four dense layers and a maximum voting classifier, This has been tested on the ASU dataset of imagined speech \cite{nguyen2017inferring}. The accuracy obtained is  comparable to the state-of-the-art results.

The rest of the paper is organized as follows: Section \ref{sec:9} describes prior work in the literature in the field of decoding imagined speech. Section \ref{sec:3} describes the dataset and  procedure for generating the feature vectors. Section \ref{sec:4} describes the classifiers in some detail. The results obtained are given in Section \ref{sec:5}.

\section{\label{sec:9} Related Work}
This section briefly describes the work in the field of imagined speech over the last decade.

C.S. DaSalla \textit{et al.} developed a BCI system based on vowel imagery \cite{dasalla2009single} in the year 2009. The objective was to discriminate between imagined vowels, \textit{/a/} and \textit{/u/}. The experimental paradigm consisted of three parts:
\begin{enumerate}
    \item Imagined mouth opening and imagined vocalisation of vowel \textit{/a/}.
    \item Imagined lip rounding and imagined vocalisation of vowel \textit{/u/}.
    \item Control state with no action.
\end{enumerate}
 
Using CSP generated spatial filter vectors as features and nonlinear SVM as a classifier, they achieved an accuracy in the range of 56\% to 72\%, depending on the subject. As noted by Brigham \textit{et.al} \cite{brigham2010imagined}, the relatively higher accuracy obtained might have arisen because of the additional involvement of motor imagery.
 
 Following a similar approach, Wang Li \textit{et al.} in 2013 developed a system to classify two monosyllabic Chinese characters meaning ``left'' and ``one'' \cite{wang2013analysis}. Visual cue was provided to the subject to instruct him/her on the character to be imagined. When the cue disappears, the subject has to repeatedly imagine the character in his/her mind as many times as possible for a duration of 4 sec. The accuracy obtained by them is around 67\%.
 
In 2010, Brigham\textit{ et al.} came up with an algorithm based on autoregressive (AR) coefficients and k-nearest neighbor (k-NN) algorithm for classifying two imagined syllables \textit{/ba/} and \textit{/ku/} \cite{brigham2010imagined}. In this experiment, the subjects were given an auditory cue on the syllable to be imagined, followed by a series of click sounds. After the last click, the subjects were instructed to imagine the syllable once every 1.5 sec for a period of 6 sec. The accuracy reported is around 61\%.

In 2016, Min \textit{et.al} used statistical features such as mean, variance, standard deviation, and skewness for pairwise classification of vowels (\textit{/a/}, \textit{/e/}, \textit{/i/}, \textit{/o/}, and \textit{/u/}) using extreme learning machine (ELM) with radial basis function. In their experimental paradigm, auditory cue  was provided at the beginning of the trial to inform the subject as to which  vowel was to be imagined. After the auditory cue, two beeps were played, after which the subject has to imagine the vowel heard during the beginning of the trial. An average accuracy of about 72\% was reported.

In 2017, Nguyen, Karavas and Artemiadis \cite{nguyen2017inferring} came up with an approach based on Riemannian manifold features for classifying four different sets of prompts:

\begin{enumerate}
\item Vowels (\textit{/a/}, \textit{/i/} and \textit{/u/}).
\item Short words (``\textit{in}'' and ``\textit{out}'').
\item Long words (``\textit{cooperate}'' and ``\textit{independent}'').
\item Short-Long words  (``\textit{in}'' and ``\textit{cooperate}'').
\end{enumerate}
The accuracy reported for the four sets of prompts are  49.2\%, 50.1\%, 66.2\% and 80.1\%, respectively. This dataset is one amongst the few imagined speech datasets that are available in the public domain and is referred to as the \textit{``ASU dataset''}. More information about this dataset is given in Section \ref{sec:2}.

Balaji \textit{et al.} in 2017  investigated the use of bilingual imaginary speech,viz., English: ``\textit{Yes''} \& ``\textit{No}''	and Hindi: ``\textit{Haan}'' (meaning ``\textit{yes}'') \& ``\textit{Na}'' (meaning ``\textit{no}'') for an imagined speech based BCI system \cite{balaji2017eeg}. PCA was used for data reduction and Artificial Neural Network (ANN) was used as the classifier. Two specific sets of EEG channels corresponding to language comprehension and decision making were utilized. An interesting part of the experimental protocol is that there is no auditory or visual cue and the subjects were instructed to imagine the answer to a binary question posed either in English or Hindi. The study reports an accuracy of 75.4\% for the combined English-Hindi task and quiet a surprising high accuracy of 85.2\% for classifying the decision.  

In 2017, Sereshkeh \textit{et al.} came up with an algorithm based on features extracted using discrete wavelet transform (DWT) and regularized neural networks for classifying the imagined decisions of ``\textit{yes}'' and ``\textit{no}'' \cite{sereshkeh2017eeg}, similar to the work by Balaji \textit{et al.} They reported an accuracy of about 67\%.

In 2018, Cooney \textit{et al.} \cite{cooney2018mel} used MFCC features and SVM classifier to classify all the 11 prompts in the KARAONE dataset \cite{zhao2015classifying}. The prompts consisted of seven phonemic/syllabic prompts (\textit{/iy/}, \textit{ /uw/},  \textit{/piy/}, \textit{/tiy/}, \textit{/diy/}, \textit{/m/}, \textit{/n/}) and four words (``pat'', ``pot'', ``knew'' and ``gnaw''). A maximum accuracy of only 33.33\% was achieved. The lower accuracy might have arisen because of a larger number of choices instead of a binary choice as in the previous works.

\section{Dataset and Methods\label{sec:3}}
\subsection{\label{sec:2} Dataset}
The ASU dataset consists of 64-channel EEG signal recorded at 1 kHz from 15 healthy subjects during four different types of imagined speech, which were described in Section \ref{sec:9}. The electrode placement was based on the 10/20 system \cite{klem1999ten}. 
	
Each trial started with a visual cue indicating the word to be imagined. Beep sound repeating at $T$ seconds was played for a period of $7\times T$ seconds after which the subject was to imagine vocalising the prompt thrice at a rate of one per $T$ seconds. For the prompt \textit{``in''}, $T$ was 1 sec and for the prompt \textit{'`cooperate''}, $T$ was 1.4 sec.

The highest accuracy reported is for the classification of a short word (\textit{``in''}) and a long word (\textit{``cooperate''}). For all participants except ``S10'' and ``S14'', EEG data of 100 trials are available for each prompt. For subjects ``S10'' and ``S14'', data of only 80 trials are available.  These two subjects are not included in our analysis due to the mis-match in the number of trials.

The EEG signals are down-sampled to 256 samples/second. A 5th order Butterworth bandpass filter with passband  from 8 to 70 Hz is applied to remove electrode drift and EMG artifacts. Further, a notch filter at 60 Hz is applied  to remove line noise artifacts.

\subsection{Channel selection}
\begin{figure*}
  \includegraphics[width=\linewidth]{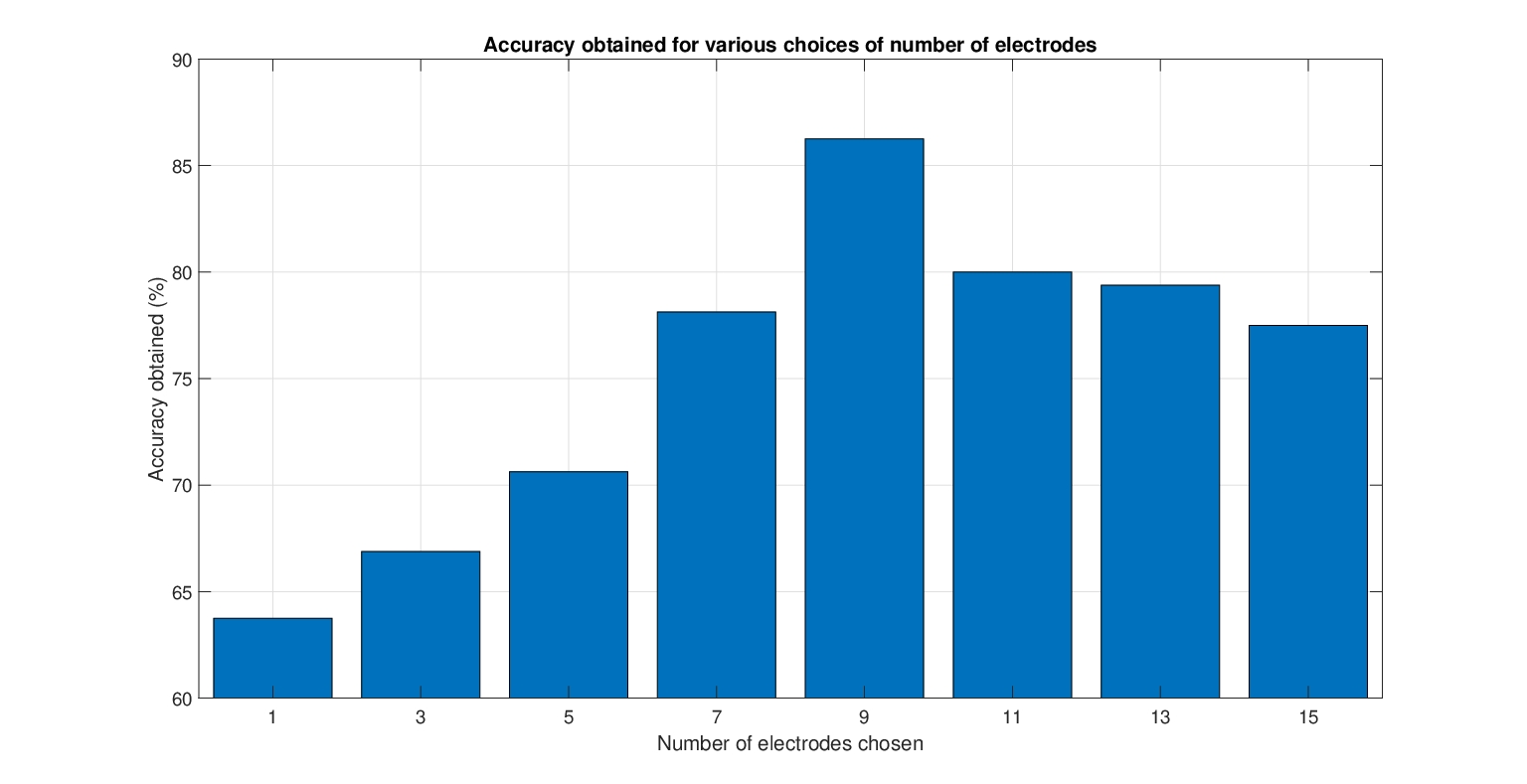}
  \caption{This figure shows the 10-fold cross-validation accuracy obtained (in percentage) for various choices of the number of electrodes for subject ``S9''. The maximum accuracy is obtained when nine channels are chosen. As evident from the figure, an increase or decrease in the number of channels reduces the accuracy. }
  \label{fig}
\end{figure*}

In our work, instead of concatenating the features obtained from several channels, each channel is viewed as a distinct input. This is possible because of the high correlation between the signals in various channels \cite{ramakrishnan2016reconstruction}. 
	
CSP linearly transforms the multi-channel EEG signal into a low-dimensional subspace such that the variance of the EEG signal from one class is maximized while that from the other class is minimized. Mathematically, CSP extremizes the following objective function:
	\begin{equation}
	J(\mathbf{w})=\frac{\mathbf{w}^TX_1X_1^T\mathbf{w}}{\mathbf{w}^TX_2X_2^T\mathbf{w}}=\frac{\mathbf{w}^TC_1\mathbf{w}}{\mathbf{w}^TC_2\mathbf{w}}
	\label{eq1}
	\end{equation}
	where $T$ denotes matrix transpose, matrix $X_i$ contains the EEG signals of class $i$, with data samples as columns and channels as rows, $\mathbf{w}$ is the spatial filter and $C_i$ is the spatial covariance matrix of  class $i$. 
	
The spatial filters can be seen as EEG source distribution vector \cite{wang2006common}. The channels corresponding to maximal coefficients in the spatial filter that maximizes $J(\mathbf{w})$ and the channels corresponding to maximal coefficients in the spatial filter that minimizes $J(\mathbf{w})$  have maximum information regarding the two classes \cite{wang2006common}.

Let $\mathbf{w}_{max}$ and $\mathbf{w}_{min}$ respectively be the spatial filter vectors that maximize and minimize the objective function $J$ given in Eq. \ref{eq1}. Since 64 channel EEG signal is used in this work, the dimension of $\mathbf{w}_{max}$ and $\mathbf{w}_{min}$ will be $64 \times 1$.

Nine channels  per class are selected corresponding to the  top nine coefficients in the vectors $\mathbf{w}_{max}$ and $\mathbf{w}_{min}$. We have observed in our experiments that the number of channels more than  or less than nine  decreases the accuracy (see Fig. \ref{fig}). The decrease when increasing the number of channels  might be because of the fact that the additional channels  may not be capturing relevant information. The decrease when the number of channels is decreased is because of the reduction in the training data. The variation in accuracy for different number of channels selected for Subject ``S9'' is given in . CSP analysis is carried out by the toolbox provided by Lotte \textit{et.al.} \cite{lotte2011regularizing}.

\subsection{Feature extraction}
	
Since each EEG channel is considered as an independent data vector, algorithms that extract a single feature vector from the entire set of EEG channels (such as Reimannian manifold features used by Nguyen \textit{et.al} \cite{nguyen2017inferring} and fuzzy entropy features \cite{raghu2018novel}) cannot be used with the proposed architecture. DWT, Daubechies-4 (db4) wavelet, is extensively used in extracting features from EEG signals \cite{nicolas2012brain}. We follow the approach of Sereshkeh \textit{et. al} in \cite{sereshkeh2017eeg}, where four levels of wavelet decomposition is performed on the input EEG signals and root-mean-square (RMS), variance and entropy are computed for each level. This yields 12 features (4 levels $\times$ 3 features) per channel.

Let $\mathbf{f}_{iA}$ be the 12 dimensional feature vector extracted from the channel corresponding to the $i^{th}$ largest coefficient in the spatial filter $\mathbf{w}_{max}$ and $\mathbf{f}_{iB}$ be the 12 dimensional feature vector extracted from the channel corresponding to the $i^{th}$ largest coefficient in the spatial filter $\mathbf{w}_{min}$. Now, $\mathbf{f}_i$, the $i^{th}$ feature vector extracted for  a specific trial is obtained by concatenating $\mathbf{f}_{iA}$ and $\mathbf{f}_{iB}$. The range of $i$ is from 1 to 9, since we are considering only nine channels corresponding to each trial. Thus, nine feature vectors of length 24 (12 from each of the two channels) are obtained per trial.

\section{Classifier\label{sec:4}}

\begin{figure}[h!]

\centering

  \includegraphics[width=11cm]{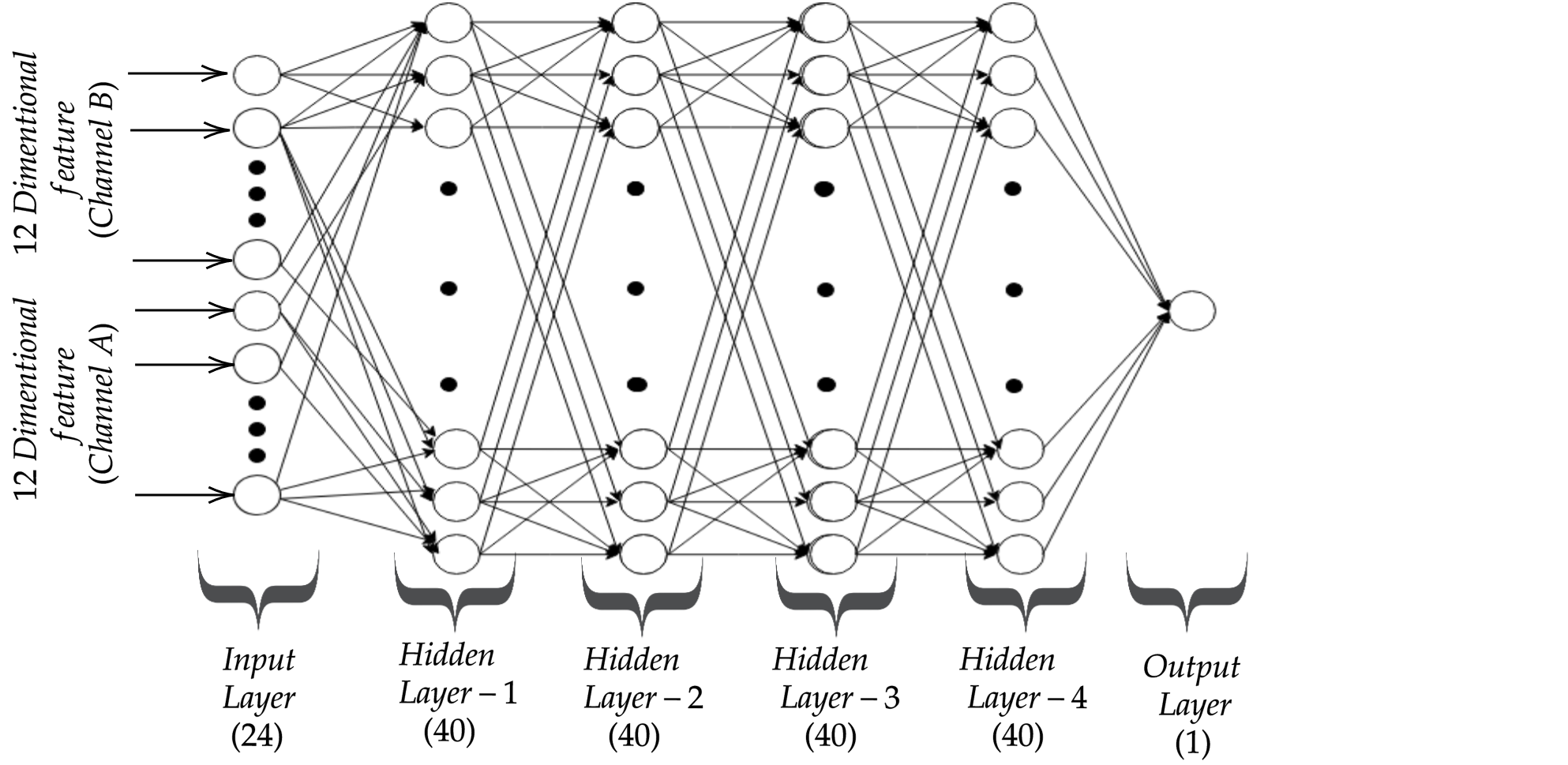}
\caption{Architecture diagram of the DNN used. The numbers in brackets indicate the number of neurons in the layer. The feature vector $i$ is obtained by concatenating the feature extracted from the channel corresponding to the $i^{th}$ largest coefficient in the spatial filter $\mathbf{w}$ that maximises $J(\mathbf{w})$ in Eq. \ref{eq1} with the features extracted from the channel corresponding to the $i^{th}$ largest coefficient in the spatial filter $\mathbf{w}$ that minimizes $J$ in Eq. \ref{eq1}. The activation function of all the hidden layers except 3 is rectified linear unit. The activation function of layer 3 is hyperbolic tangent. The activation function of the single output neuron is \textit{sigmoid}. Dropout and batch normalization layers are not shown in the diagram.}
	\label{dnn}
\end{figure}

A DNN with four hidden layers is used as the primary classifier. The architecture of the DNN used is shown in Fig. \ref{dnn}. Each dense hidden layer has 40 neurons. Also, dropout and batch normalization layers are added after each dense layer. The dropout ratio is 10\% for the dropout layer after the first hidden layer and all the other dropout layers has a dropout ratio of 30\%. The activation function of all the hidden layers except third is rectified linear unit. The activation function of layer 3 is hyperbolic tangent. The activation function of the single output neuron is \textit{sigmoid}. 

Since we have derived nine feature vectors per trial, nine outputs are obtained for each trial, one for each feature vector. The final decision is based on majority or hard voting.

{\def\arraystretch{2}\tabcolsep=10pt
	\begin{table*}[h!]
	\centering
	\caption{Comparison of mean $\pm$ std. deviation of the cross-validation accuracies in percentage obtained using different methods (given in each row) in classifying imagined words, ``\textit{in}'' and ``\textit{cooperate}''. ``S1'', ``S5'', ``S8'' and ```S9'' are the participant IDs. }
\begin{tabular}{|l|c|c|c|c|}
	\hline
	\multicolumn{1}{|c|}{Method/Subject}                                        & \textbf{S1}     & \textbf{S5}     & \textbf{S8}     & \textbf{S9}     \\ \hline
	\textbf{Log + LDA}  \cite{nguyen2017inferring}                                                        & $50.5\pm 14.8$  & $59.5 \pm 5.7$  & $36.9\pm 15.9$  & $74.1 \pm 16.6$ \\ \hline
	\textbf{CSP + SVM} \cite{dasalla2009single}                           & $61.5 \pm 12.0$ & $61.5 \pm 8.8$  & $62.5 \pm 8.3$  & $58.1 \pm 7.2$  \\ \hline
	\textbf{Statistical features + ELM} \cite{min2016vowel}                               & $51.0 \pm 8.4$  & $59.5 \pm 6.4$  & $59.4 \pm 11.5$ & $51.9 \pm 6.6$  \\ \hline
	{\begin{tabular}[c]{@{}l@{}}\textbf{Tangent + RVM} \textbf{ (Method 1) }\cite{nguyen2017inferring} \end{tabular}} & $63.3 \pm 2.9$  & $65.8 \pm 3.1$  & $76.9 \pm 3.0$  & $69.4 \pm 7.5$  \\ \hline
	{\begin{tabular}[c]{@{}l@{}}\textbf{Tangent + RVM} \textbf{ (Method 2) }\cite{nguyen2017inferring} \end{tabular}}  & $70.3 \pm 5.5$  & $71.5  \pm 5.0$ & $81.9 \pm 6.5$  & $88.0 \pm 6.4$  \\ \hline
	\textbf{Proposed Method}                                                    & $65.5 \pm 9.6$  & $64.5 \pm 10.3$ & $71.0 \pm 5.3$  & $86.2 \pm 8.7$  \\ \hline
\end{tabular}
\label{t1}
	\end{table*}
	
\section{Results\label{sec:5}}

10-fold cross-validation is performed on the pre-processed data of each participant. During cross-validation, it is ensured that all the channels corresponding to a trial are either in the training set or in the test set. This is important, since the presence of a couple of channels from the test trials in the training set can lead to high spurious accuracy due to data leakage. The results obtained along with the other results reported in the literature are listed in Table \ref{t1}. The proposed method gives a higher accuracy compared to other methods except ``Tangent + RVM Method (2)'' which has around 2 to 10\% higher accuracy. The average accuracy obtained across the participants is $71.8 \pm 8.6 \%$ and the best accuracy reported in the literature is $77.6 \pm 5.7 \%$. Although the accuracy of the proposed method is lower than the best reported in the literature, the motivation for the present study is to explore the possibility of using DNN in the context of imagined speech. Hopefully, the accuracy may be improved further with denser EEG acquisition system.

\section{\label{sec:l} Conclusion}

The present work shows that it is feasible to view each EEG channel as an independent data vector in order to increase the size of the training set for the purpose of classification  in decoding imagined speech using deep learning techniques. We expect an  improved performance by acquiring EEG data using higher density EEG acquisition systems. For instance, using a 128-channel EEG system instead of the 64-channel ASU dataset will give more amount of data for training for the same number of trials. This can in turn help us using a more complex network for classification.

\bibliographystyle{IEEEtran}
\bibliography{mile3}
\end{document}